\documentclass[aps,preprint,showpacs,amsmath,amssymb]{revtex4}
\usepackage{epsfig}
\usepackage{wasysym}
\usepackage{amssymb}

\begin{document}

\newcommand{\bvec}[1]{\mbox{\boldmath ${#1}$}}
\title{Photo- and electroproduction of the $K^0\Lambda$ near 
  threshold and effects of the $K^0$ electromagnetic form factor}
\author{T. Mart}
\affiliation{Departemen Fisika, FMIPA, Universitas Indonesia, Depok 16424, 
  Indonesia}
\date{\today}
\begin{abstract}
By extending our previous isobar model to the 
$K^0\Lambda$ isospin channel we investigate 
the properties of the $K^0\Lambda$ photo- and 
electroproduction at energies near threshold. 
It is found that the pseudovector (PV) coupling yields
significantly larger cross section. Variation of the $K_1$
coupling constants has significant effect only on the PV model.
The electromagnetic 
form factor of the neutral kaon  $K^0$ is found to have a 
sizable effect on the longitudinal cross section of the $K^0\Lambda$
electroproduction near the threshold.
\end{abstract}

\maketitle

Recently, there has been a great interest in the 
electromagnetic production of the neutral kaon 
\cite{Salam:2006kk,motivation}. 
There are several motivations behind this interest,
one of them is the search for missing resonances.
In the $K^0$ photoproduction the $t$-channel $K^0$
intermediate state does not present 
due to the lack of the interaction between real photons and
neutral mesons. The absence of this channel might 
have a strong impact on the angular distribution of the 
predicted observables and also increase the
dominance of the nucleon resonance contribution. 
This situation is obviously different in 
the leading $K^+\Lambda$ \cite{missing-d13} and 
$K^+\Sigma^0$ \cite{Mart:1995wu} channels, which for
years have become the main source of information on the 
strangeness production process. Another motivation
comes from the deuteron sector, in which the relevant
processes are the $\gamma + d\to K^0+\Lambda+p$ and 
$\gamma + d\to K^0+\Sigma^0+p$. Due to the lack of
the neutron target, the two processes are expected
to be the natural avenue in the investigation of
kaon photoproduction on the neutron. A recent report
shows that the extraction of the elementary cross 
section is possible for these processes in the
quasifree scattering region, where the final state
interaction effects are negligible \cite{Salam:2006kk}.
Note that the result of such an analysis relies on the 
result of Kaon-Maid model \cite{kaon-maid}, especially 
on the $K^0\Lambda$ channel. However, the predicted
total cross section of this channel is found to be
twice larger than that of the $K^+\Lambda$ channel,
in contrast to the case of other four related isospin channels
(see Fig. 1 of Ref.~\cite{Salam:2006kk}).

There was an attempt to investigate the effect of the
$K^0$ charge form factor on the $K^0\Lambda$
electroproduction \cite{Mart:1997cc}. Although  
significant effects on the longitudinal cross section 
were observed, the result was found to be 
very model dependent. Consequently, a reliable 
phenomenological model becomes the important
prescription for this purpose.
Since a reliable model should not depend on too 
many uncertain free parameters, it is obviously 
important to limit the energy of interest 
very close to the production threshold.

Very recently, we have analyzed photo- and
electroproduction of the $K^+\Lambda$ final state 
at energies near its production threshold by utilizing 
an isobar model \cite{Mart:2010ch}. Using 
the pseudoscalar (PS) coupling the model can nicely 
describe experimental data both in the real and virtual 
photon sectors up to total c.m. energy $W=50$ MeV above 
the threshold. Using the pseudovector (PV) coupling the 
agreement with experimental data can still be achieved, 
although the $\chi^2$ per number of data points increases 
from 0.92 to 1.53. 

This paper reports on the extension of our 
previous model  
to the $K^0\Lambda$ isospin channel. 
For this purpose we employ the SU(3) 
symmetry to relate the hadronic coupling constants of the
background terms in the two channels, i.e.,
\begin{equation}
\label{eq:coupling1}
g_{K^{+} \Lambda p} = g_{K^{0} \Lambda n} ,~
g_{K^{+} \Sigma^0 p} = -g_{K^{0} \Sigma^0 n} ,~
g^{V,T}_{K^{*+} \Lambda p} = g^{V,T}_{K^{*0} \Lambda n} .
\end{equation}
In the $K^{0}\Lambda$ production the vector meson exchanged 
in the $t$-channel is the $K^{*0} (896.10)$. Therefore, the 
transition moment in $K^{+}$ production, $g_{K^{+*}K^+ \gamma}$,  
must be replaced by the neutral transition moment by using 
\cite{Mart:1995wu}
\begin{eqnarray}
g_{K^{*0} K^{0} \gamma}/g_{K^{*+} K^{+} \gamma} = -1.53\pm 0.20 \,.
\end{eqnarray}
In the case of the $K_1(1270)$ vector meson exchange, there is
no sufficient information from the Particle Data Book \cite{pdg2010}. 
Thus, we use the value given by the Kaon-Maid \cite{kaon-maid},
\begin{eqnarray}
r_{K_1K\gamma}\equiv
g_{K_1^{0} K^{0} \gamma}/g_{K_1^{+} K^{+} \gamma} = -0.45 \,,
\label{eq:ratio_K1}
\end{eqnarray}
which was extracted from 
simultaneous fitting of the $K^+\Sigma^0$ and
$K^0\Sigma^+$ photoproduction data. The 
$S_{01}(1800)$ $u$-channel stays unmodified, since 
from Eq.~(\ref{eq:coupling1}) we have
$g_{K^{+} \Lambda p} = g_{K^{0} \Lambda n}$, whereas the electromagnetic
vertex $\gamma Y^{*0}\Lambda$
in both $K^+\Lambda$ and $K^0\Lambda$ channels
is the same.

\begin{figure}[t]
  \begin{center}
    \leavevmode
    \epsfig{figure=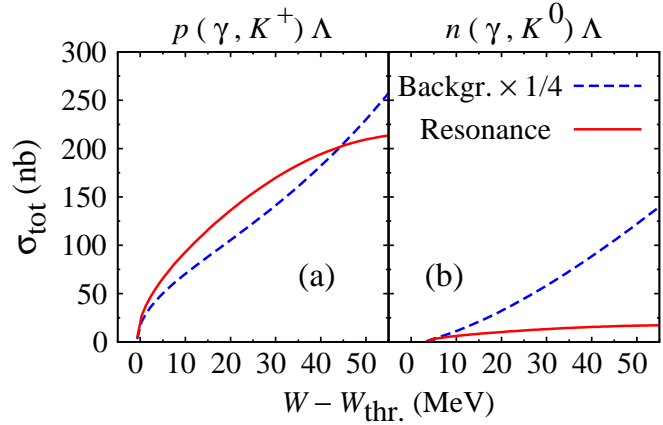,width=90mm}
    \caption{(Color online) Contributions of the background terms and the 
      $S_{11}(1650)$ resonance to the total cross section 
      of the $\gamma+p\to K^++\Lambda$ (a)
      and $\gamma+n\to K^0+\Lambda$ (b) channels.}
   \label{fig:contrib} 
  \end{center}
\end{figure}

In the resonance term we have to replace
the helicity photon couplings of the proton
$A_{1/2}^p$ with that of the neutron, where
\cite{pdg2010}
\begin{eqnarray}
\label{eq:coupling2}
A_{1/2}^n = -0.015\pm 0.021 ~{\rm GeV}^{-1/2} \,.
\end{eqnarray}
Note that this coupling is substantially smaller
than the proton coupling, i.e.
$A_{1/2}^p=0.053\pm 0.016$ GeV$^{-1/2}$ \cite{pdg2010}. 
As a consequence, we may expect a relatively smaller
resonance contribution in the case of $K^0\Lambda$
production. This is proven by Fig.~\ref{fig:contrib},
where we compare the background and resonance contributions
to the total cross section of the $K^+\Lambda$ and $K^0\Lambda$
photoproduction. Different from the $K^+\Lambda$ channel,
in which contribution of the $S_{11}(1650)$ resonance is
more or less 20\%, contribution of this resonance to the
$K^0\Lambda$ total cross section at $W=50$ MeV above 
threshold is only about 3\%.

\begin{figure}[b]
  \begin{center}
    \leavevmode
    \epsfig{figure=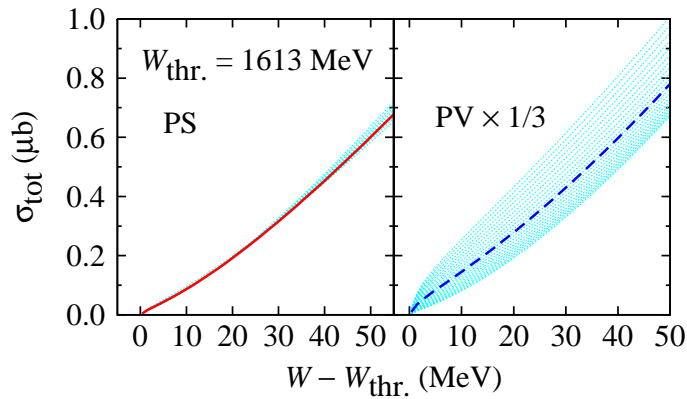,width=90mm}
    \caption{(Color online) Total cross sections 
      of the $\gamma+n\to K^0+\Lambda$ channel predicted by the models
      using PS (left panel) and PV (right panel) couplings. The shaded
      area corresponds to the variation of the
      the ratio $r_{K_1K\gamma}$ in
      Eq.~(\ref{eq:ratio_K1}). The PV cross section has been renormalized
      by a factor of 1/3.}
   \label{fig:total} 
  \end{center}
\end{figure}

The predicted $K^0\Lambda$ total cross sections of both 
PS and PV models are shown in Fig.~\ref{fig:total}, where 
the effects of the variation of the ratio $r_{K_1K\gamma}$ given in 
Eq.~(\ref{eq:ratio_K1}) are displayed. For the sake of visibility,
we have varied this ratio by $\pm 100$\% ($\pm 20$\%) in the PS (PV)
model. Both the cross section and the effect are obviously 
larger in the PV model. We have found that this phenomenon 
originates from the large $K_1$ and $K^*$ couplings
in the PV model. Especially in the case of $K_1$, where
the coupling constants are around 10 times larger than
those in the PS model.
Although we believe that the PS model is still better than the
PV one, as in the $K^+\Lambda$ case, an experimental check of the 
$K^0\Lambda$ total cross section is still mandatory to  
help to clarify this situation.

\begin{figure}[t]
  \begin{center}
    \leavevmode
    \epsfig{figure=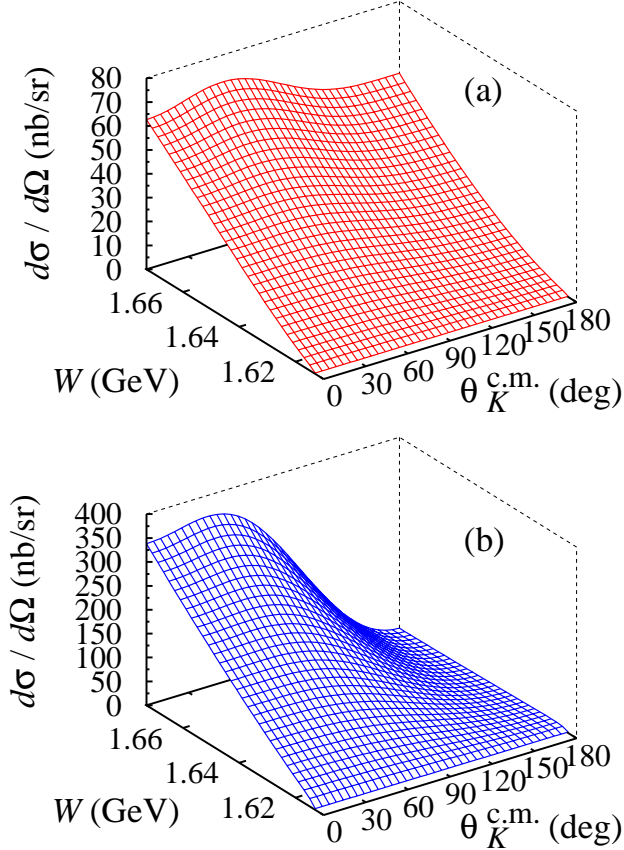,width=85mm}
    \caption{(Color online) Differential cross sections of the 
      $\gamma+n\to K^0+\Lambda$ channel predicted by the models
      using pseudoscalar (a) and pseudovector (b) couplings.}
   \label{fig:diff} 
  \end{center}
\end{figure}

The difference between PS and PV models also appears in the 
differential cross section as shown 
in Fig.~\ref{fig:diff}. From this figure
it is apparent that the dominant role of 
the $K_1$ and $K^*$ exchanges in the PV
model yields not only a large cross section, but also
amplifies the bump structure in the angular distribution
of differential cross section. 
Therefore, experimental data of the $K^0\Lambda$
differential cross section can shed more light on
the role of the $K_1$ and $K^*$ in kaon photoproduction.
From Fig.~\ref{fig:diff} we can see that both models 
do not indicate a backward-peaking cross section.
This is different from the cross section estimated by 
the deuteron target \cite{motivation}.
However, we realize that in order to estimate  this 
cross section the ratio $r_{K_1K\gamma}$ is varied 
\cite{motivation}. It is important to note here  
that this ratio is no longer a free parameter if one starts
with the elementary process, but it is fixed by 
the $K^+\Sigma^0$ and $K^0\Sigma^+$ photoproduction 
channels, for which more experimental data with better 
statistics are available. Changing the value of $r_{K_1K\gamma}$ 
will obviously change the predicted 
observables in the  $K^0\Sigma^+$ channel. 
Furthermore, we also note that the elementary amplitude used to
extract the cross section (called SLA in \cite{motivation}) 
fits relatively older data. In the previous work \cite{Mart:2010ch}
we used more recent data and 
found that the new electroproduction data provide a stringent
constraint to the background, especially to the $K_1$ contribution.

The extension of the PS model to the case of 
electroproduction has been also discussed in our 
previous report \cite{Mart:2010ch}. Fortunately, experimental
data are also available for the $K^+\Lambda$ electroproduction
near threshold so that all unknown longitudinal/scalar 
couplings can be directly extracted. However, this is not the 
case in the $K^0\Lambda$ electroproduction. Thus, the required 
scalar photon coupling of the $S_{11}(1650)$ resonance 
amplitude is taken from the MAID2007 model 
\cite{Drechsel:2007if}, i.e.
$S_{1/2}^n = 0.010 ~{\rm GeV}^{-1/2}$.
The neutron electromagnetic form factors are taken from the 
Galster parameterization \cite{Galster:1971kv}, whereas the 
hyperon form factors as well as the dependencies of the electric 
and scalar multipoles on the $Q^2$ are assumed to have 
the same forms as in the $K^+\Lambda$ channel \cite{Mart:2010ch}. 

Compared with other neutral SU(3) pseudoscalar mesons, 
the neutral kaon has a unique property, i.e. it has
an electric or charge form factor. The difference between the strange 
and non-strange quark masses creates a non-uniform charge 
distribution in the $K^0$. Consequently, although its
total charge is zero, the $K^0$ has an electric or charge form factor. 
Since the mass difference is still smaller than the mass scale 
associated with confinement in Quantum Chromodynamics (QCD), 
$(m_s-m_d)<\Lambda _{\rm QCD}$, it could lead to a sensitive 
test of phenomenological models that attempt to describe 
nonperturbative QCD.

\begin{figure}[t]
  \begin{center}
    \leavevmode
    \epsfig{figure=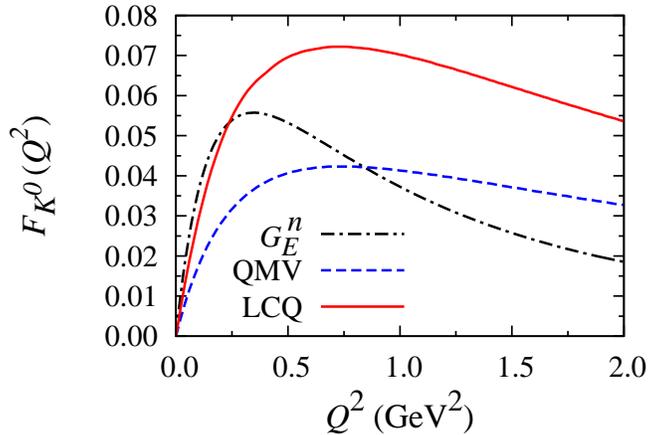,width=85mm}
    \caption{(Color online) Electromagnetic form factors of the
    neutral kaon predicted by the Quark Meson Vertex (QMV) and
    Light Cone Quark (LCQ) models compared with the electric
    form factor of the neutron $G_E^n$.}
   \label{fig:ff_k0} 
  \end{center}
\end{figure}

In this paper we do not intent to discuss the form factor
in details, instead we will only employ two relativistic 
quark models, the light-cone quark (LCQ) model~\cite{ito1} 
and the quark-meson vertex (QMV) model~\cite{buck}, in order
to find the optimal kinematics where this form factor has
the largest effect on the observable. Since contributions
of the $K^*$ and $K_1$ exchanges are relatively small in
the PS model, contribution of the kaon pole is expected 
to generate sizable effects on the cross section.
The characteristic of these 
two form factors is exhibited 
in Fig.~\ref{fig:ff_k0}, where the charge form 
factor of the neutron $G_E^n(Q^2)$ 
\cite{Galster:1971kv} is also shown 
for comparison. It is clear from this figure that they
have comparable magnitudes. The only difference is that 
as $Q^2$ increases the two neutral kaon form factors 
fall off slower than the neutron one. 

We note that the inclusion of the $K^0$ form factor is
connected with the problem of gauge invariance. In this
work we have utilized the Fubini-Nambu-Wataghin term 
\cite{fnw} to restore gauge invariance, which is discussed 
in Ref.~\cite{deo}. Since the term is very specific and
not trivial, we expect that different methods of restoring 
gauge invariance in this process will not affect the result
shown in this work.

\begin{figure}[t]
  \begin{center}
    \leavevmode
    \epsfig{figure=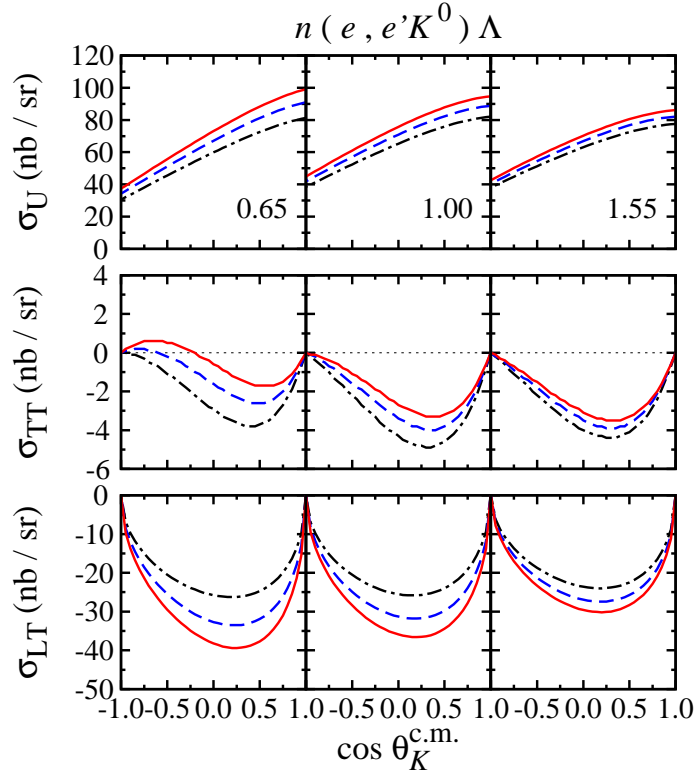,width=95mm}
    \caption{(Color online) Separated differential cross sections for the 
      neutral kaon electroproduction $e+n\to e'+K^0+\Lambda$ as a function
      of the kaon scattering angle at $W=1.65$ GeV and for different 
      values of $Q^2$ (shown in the top panels in unit of GeV$^2$). 
      Solid lines show the calculation with a $K^0$ form factor obtained 
      in the LCQ model while dashed lines are obtained by using the 
      QMV model. The dash-dotted lines are obtained from a computation with 
      the $K^0$ pole excluded. Note that $\sigma_U = d\sigma_T/d\Omega_K + 
      \epsilon d\sigma_L/d\Omega_K$, $\sigma_{TT} = d\sigma_{TT}/d\Omega_K$, 
      and $\sigma_{LT} = d\sigma_{LT}/d\Omega_K$.}
   \label{fig:ambroz} 
  \end{center}
\end{figure}

The effect of the QMV and LCQ form factors on the separated
differential cross sections of the $K^0\Lambda$ electroproduction on 
a neutron is displayed in Fig.~\ref{fig:ambroz}, where we have
chosen a closer kinematics as in the case of the $K^+\Lambda$
\cite{Mart:2010ch}, since we expect that with the present
technology such a kinematics is experimentally accessible.
From this figure it is apparent that the observed effect in
the cross section magnitude is consistent with the behavior
of the form factors exhibited in Fig.~\ref{fig:ff_k0}, i.e.
the LCQ model yields the strongest effect.

\begin{figure*}[t]
  \begin{center}
    \leavevmode
    \epsfig{figure=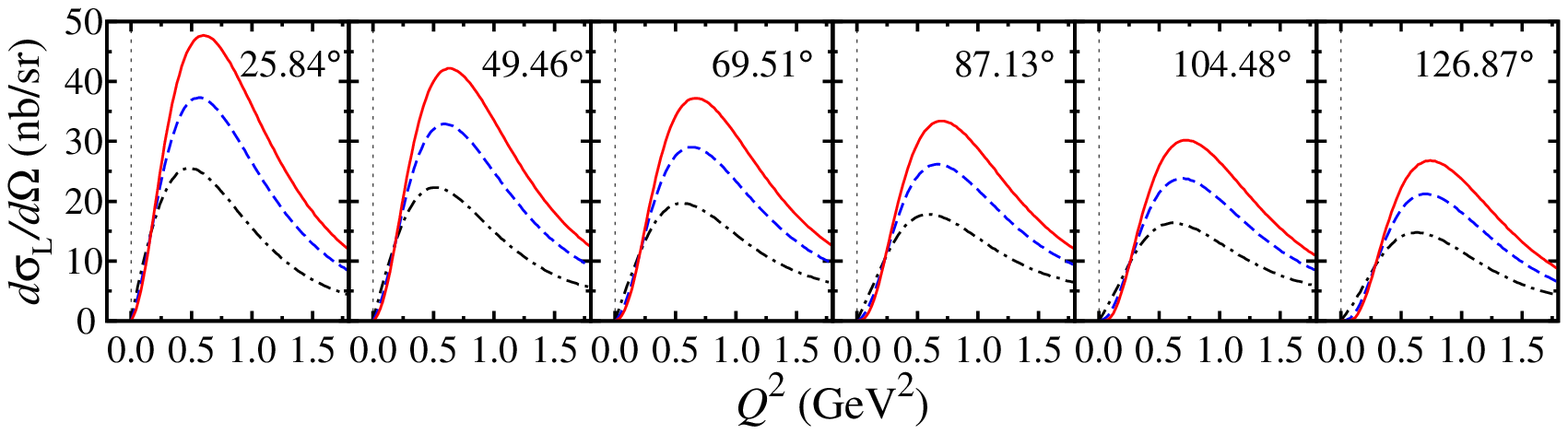,width=160mm}
    \caption{(Color online) Longitudinal differential cross section of the 
      neutral kaon electroproduction $e+n\to e'+K^0+\Lambda$ as a function
      of the virtual photon momentum squared $Q^2$ 
      at $W=1.65$ GeV and for different values of the 
      kaon scattering angles (shown inside 
      the panels). Notation of the curves is as in 
      Fig.~\ref{fig:ambroz}.}
   \label{fig:sigLsigT} 
  \end{center}
\end{figure*}

The effect of $K^0$ form factors on the unpolarized cross section
$\sigma_{\rm U}$ shown in the upper panels of Fig.\,\ref{fig:ambroz}
seems to be mild, in contrast to the effect on the separated cross
sections $\sigma_{\rm TT}$ and $\sigma_{\rm LT}$.
Especially in the case of the interference cross section
 $\sigma_{\rm LT}$ shown in the bottom
panels of Fig.\,\ref{fig:ambroz}, where the effect of the
LCQ model is predicted to be around 30\% at $Q^2=0.65$ GeV$^2$.
However, we found that the effect is almost
negligible on the transversely unpolarized cross section 
$d\sigma_{\rm T}/d\Omega$. Therefore, the difference
between the three lines shown in this figure 
originates mostly from the longitudinal
cross section $d\sigma_{\rm L}/d\Omega$. 
Consequently, we then focus our analysis 
on the longitudinal cross section.
The effects for different values of kaon scattering
angle are shown in Fig.~\ref{fig:sigLsigT}. It is obvious that the
effect is sufficiently large for an experimental check and 
in fact at the forward 
angle $\theta_K^{\rm c.m.}=25.84^\circ$ and $Q^2\approx 0.5$ GeV$^2$
the LCQ form factor raises the cross section up to 50\%. 
As explained above, this phenomenon originates from the dominant 
role of the background terms. From this figure it is also clear 
that as the scattering angle 
increases the effect slightly decreases but still relatively 
large for the LCQ model at  $\theta_K^{\rm c.m.}=126.87^\circ$. 
Our finding therefore corroborates the finding of 
Ref.\,\cite{Mart:1997cc}, which used the same form factors
\cite{ito1,buck} as in the present work but with a different 
isobar model~\cite{williams}. Experimental data with about
10\% uncertainties would be able to resolve the effect of
the form factors or even to pin down the appropriate $K^0$ form
factor required by the isobar model to describe the $e+n\to e'+K^0+\Lambda$
process.

In conclusion we have extended our previous isobar model for the
$K^+\Lambda$ channel to include both photo- and electroproduction 
of the $K^0\Lambda$ by exploiting the SU(3) symmetry and 
appropriate information from the Particle Data Book. 
Using this model we have also analyzed the differences 
in the total and differential cross sections obtained by using PS and PV 
couplings. It is found that the PV model yields significantly larger 
total and differential cross sections. 
Needless to say, experimental data on the 
$K^0\Lambda$ photoproduction are required to check this 
phenomenon.
We have also used the PS model to explore the effect
of the $K^0$ charge form factor and found sizable effects on
the longitudinal and interference cross sections of the $K^0\Lambda$ 
electroproduction. Especially in the longitudinal cross section,
we found that the effect could raise the cross section up to
50\%.

\section*{Acknowledgment}
The author acknowledges supports from the University of Indonesia
and the Competence Grant of the Indonesian 
Ministry of National Education.


\begin{thebibliography}{0}
\bibitem{Salam:2006kk}
  A.~Salam, K.~Miyagawa, T.~Mart, C.~Bennhold and W.~Gl\"ockle,
  Phys.\ Rev.\  C {\bf 74}, 044004 (2006).
\bibitem{motivation} K.~Tsukada {\it et al.}, Phys.\ Rev.\  
  C {\bf 78}, 014001 (2008).
\bibitem{missing-d13} T.~Mart and C.~Bennhold, 
  Phys.\ Rev.\ C {\bf 61}, 012201 (1999).
\bibitem{Mart:1995wu}
  T.~Mart, C.~Bennhold and C.~E.~Hyde-Wright,
  Phys.\ Rev.\  C {\bf 51}, 1074 (1995).
\bibitem{kaon-maid} Available at the Maid homepage 
  http://www.kph.uni-mainz.de/MAID/kaon/kaonmaid.html. 
  The published versions can be found in: Ref. \cite{missing-d13}; 
  T.~Mart, Phys.\ Rev.\ C {\bf 62}, 038201 (2000); C.~Bennhold,
  H.~Haberzettl and T.~Mart, arXiv:nucl-th/9909022.
\bibitem{Mart:1997cc}
  T.~Mart and C.~Bennhold, Nucl.\ Phys.\  A {\bf 639}, 237 (1998).
\bibitem{Mart:2010ch}
  T.~Mart, Phys.\ Rev.\  C {\bf 82}, 025209 (2010).
\bibitem{pdg2010} K. Nakamura {\it et al.}, 
  J. Phys. G {\bf 37}, 075021 (2010).
\bibitem{Mart:2006dk}
  T.~Mart and A.~Sulaksono,
  Phys.\ Rev.\ C {\bf 74}, 055203 (2006).
\bibitem{Drechsel:2007if}
  D.~Drechsel, S.~S.~Kamalov and L.~Tiator,
  Eur.\ Phys.\ J.\  A {\bf 34}, 69 (2007).
\bibitem{ito1} C. Bennhold, H. Ito, and T. Mart, {\it Proceedings of the
               7th International Conference on the Structure of Baryons},
               Santa Fe, New Mexico, 1995, p.323.
\bibitem{buck}  W. W. Buck, R. Williams, and H. Ito, Phys. Lett. B 
  {\bf 351}, 24 (1995); H.~Ito and F. Gross, Phys. Rev. Lett. {\bf 71},
  2555 (1993).
\bibitem{Galster:1971kv} S.~Galster, H.~Klein, J.~Moritz, K.~H.~Schmidt, 
  D.~Wegener and J.~Bleckwenn, Nucl.\ Phys.\  B {\bf 32}, 221 (1971).
\bibitem{fnw} S. Fubini, Y. Nambu, and V. Wataghin, Phys. Rev. {\bf 111},
    329 (1958).
\bibitem{deo} B. B. Deo and A. K. Bisoi, Phys. Rev. D {\bf 9}, 288 (1974).
\bibitem{williams} R. A. Williams, C.-R. Ji, and S. R. Cotanch, 
                   Phys. Rev. C {\bf 46}, 1617 (1992).
\end{thebibliography}
\end{document}